\documentclass[aps,12pt,showpacs]{revtex4}

\usepackage{color,amsmath,amssymb}
\usepackage{graphicx}

\definecolor{violet}{rgb}{0.7,0,0.7}
\definecolor{lila}{rgb}{1,0,1}

\begin{document}

\title{Lower and upper bounds on the fidelity susceptibility}

\author{ J.G. Brankov$^{\dag^1\dag^2}$ and N.S. Tonchev$^{\dag^3}$
\email{brankov@theor.jinr.ru}}

\affiliation{$^{\dag^1}$~Bogoliubov Laboratory of Theoretical
Physics, Joint Institute for Nuclear Research, 141980 Dubna, Russian Federation}

\affiliation{$^{\dag^2}$~Institute of Mechanics, Bulgarian Academy
of Sciences, 4 Acad. G. Bonchev St., 1113 Sofia, Bulgaria}

\affiliation{$^{\dag^3}$~Institute of Solid State Physics, Bulgarian Academy of Sciences,
72 Tzarigradsko Chaussee Blvd., 1784 Sofia, Bulgaria}

\begin{abstract}

We derive upper and lower bounds on the fidelity susceptibility in terms of macroscopic thermodynamical quantities,
like susceptibilities and thermal average values. The quality of the bounds is checked by the exact expressions for a
single spin in an external magnetic field. Their usefulness is illustrated by two examples of many-particle
models which are exactly solved in the thermodynamic limit:
the Dicke superradiance model and the single impurity Kondo model. It is shown that as far as
divergent behavior is considered, the fidelity susceptibility and the thermodynamic susceptibility
are equivalent for a large class of models exhibiting critical behavior.

\end{abstract}

\maketitle

\section{Introduction}

Over the last decade there have been impressive theoretical advances concerning
the concepts of entanglement and fidelity from quantum and information theory
\cite{NCh00},\cite{BZ06},\cite{AFOV08}, and  their application in condensed matter physics, especially in
the theory of critical phenomena and phase transitions, for a review see \cite{AF09,Gu10}.
These two concepts are closely related to each other.

The entanglement  measures the strength of non-local quantum correlations between partitions of a compound system.
So it is natural  to expect  that entanglement will be a reliable indicator of a quantum critical point
driven  by quantum fluctuations. To this end the main efforts have been focused on the different entanglement measures
and their behavior in the vicinity of the critical point \cite{AFOV08,AF09,Gu10}.

The  fidelity \cite{U76,W81,J94,Sh95} naturally appears in quantum mechanics as the
absolute value of the overlap (Hilbert-space scalar product) of two quantum states corresponding to different
values of the control parameters.
 The corresponding finite-temperature extension, defined as a
functional of two density matrices, $\rho_1$ and $\rho_2$,
\begin{equation}
{\cal F}(\rho_1,\rho_2) = \mathrm{Tr}\sqrt{\rho_1^{1/2} \rho_2 \rho_1^{1/2}}, \label{defFidel}
\end{equation}
was introduced by Uhlmann  \cite{U76} and called fidelity by Jozsa \cite {J94}. Actually,
the definition given by Jozsa, and used, e.g., by the authors of \cite{PM09,MPHUZ,PM11,SZ03}, is the square of
${\cal F}(\rho_1,\rho_2)$, however we shall adhere to the expression (\ref{defFidel}), as most authors do.
This functional has become an issue of extensive investigations \cite{ZVG07,ZQWS07,QC09,V10,AASC10}.

Closely related to the Uhlmann fidelity (\ref{defFidel}) is the Bures distance \cite{B69},
\begin{equation}
d_{B}(\rho_1,\rho_2)=\sqrt{2-2{\cal F}(\rho_1,\rho_2)}
\label{Bures}
\end{equation}
which is a measure of the statistical distance between two density matrices. The Bures distance has the important
properties of being Riemannian and monotone metric on the space of density matrices.

Being a measure of the similarity between quantum states, both pure or mixed, fidelity should decrease abruptly at a critical
point, thus locating and characterizing the phase transition. Different finite-size scaling behaviors of the fidelity indicate
different types of phase transitions \cite{ZB08}.
The fidelity approach is basically a metric one \cite{ZQWS07}  and has an advantage over the traditional
Landau-Ginzburg theory, because it avoids possible difficulties in identifying the notions of order parameter, symmetry breaking,
 correlation length and so it is suitable for the study of different kinds of topological  or
Berezinskii-Kosterlitz-Thouless  phase transitions.

Due to the geometric meaning of fidelity, the problem of similarity (closeness) between states can be readily
translated in the language of information geometry \cite{W81,ZGC07}. The strategy here is to make an identification
of Hilbert-space geometry with the information-space geometry. Note that fidelity depends on two density matrices, $\rho_1$
and $\rho_2$, i.e. on the corresponding two points on the manifold of density matrices. On the other hand, being
sensitive to the dissimilarity of states, it could be used to measure the loss of information encoded in quantum states.

The above mentioned decrease in the fidelity ${\cal F}(\rho_{1}, \rho_{2})$, when the state $\rho_{2}$ approaches a quantum
critical state $\rho_{1}$, is associated with a divergence of the fidelity susceptibility $\chi_F(\rho_{1})$
\cite{YLG07} which reflects the
singularity of ${\cal F}(\rho_{1}, \rho_{2})$ at that point.
The fidelity susceptibility $\chi_F(\rho_{1})$, which is the main objects of this study,
naturally arises as a leading-order term
in the expansion of the fidelity for two infinitesimally close density matrices $\rho_{1}$ and $\rho_{2}=
\rho_{1} + \delta \rho$. For simplicity, in our study we consider one-parameter family of Gibbs states
\begin{equation}
\rho(h) = [Z(h)]^{-1}\exp[-\beta H(h)], \label{roh}
\end{equation}
defined on the family of Hamiltonians of the form
\begin{equation}
H(h) = T - h S, \label{ham}
\end{equation}
where the Hermitian operators $T$ and $S$ do not commute in the
general case, $h$ is a real parameter, and $Z(h)= {\mathrm Tr}\exp[-\beta H(h)]$
is the corresponding partition function. Note that the fidelity and fidelity susceptibility under consideration are
defined with respect to the parameter $h$, including the important symmetry breaking case when the system undergoes
a phase transition as $h$ is varied. The fidelity susceptibility at the
point $h=0$ in the parameter space is defined as (see e.g.\cite{V10}):
\begin{equation}
\chi_{F}(\rho(0)):=\lim _{h \rightarrow 0}\frac{-2\ln {\cal F}(\rho(0),\rho(h))}{h^{2}}=
-\left. \frac{\partial^{2}{\cal F}(\rho(0),\rho(h))}{\partial h^{2}}\right|_{h=0}  .
\label{dchi}
\end{equation}

From (\ref{Bures}) and (\ref{dchi}) we obtain for the case of two infinitesimally close density matrices the
following relation between the Bures distance and the fidelity susceptibility:
\begin{equation}
d^{2}_{B}(\rho(0),\rho(h))=\chi_{F}(\rho(0))h^{2} +O(h^{4}),\qquad h \rightarrow 0.
\label{infBures}
\end{equation}

To avoid confusion, we warn the reader that for mixed states the definition of the fidelity susceptibility
(\ref{dchi}), based on the Uhlmann fidelity (\ref{defFidel}), differs from the one derived in \cite{AASC10}
(see also \cite{Gu10}) by extending the ground-state
Green's function representation to nonzero temperatures. This fact has been pointed out in \cite{S10},
see also our Discussion.

The quantity (\ref{dchi}) is more convenient for studying than the fidelity itself,
because it depends on a single point $h=0$ and does not depend on the difference in the
parameters of the two quantum states \cite{V10,SLYM09}.
Physically, it is a measure of the fluctuations of the driving term introduced
in the Hamiltonian through the parameter $h$. In the case when $\rho_1$ and $\rho_2$ commute, there are
simple relations between the fidelity susceptibility (\ref{dchi}) and thermal fluctuations. For examples, in the case of two
states at infinitesimally close temperatures the fidelity susceptibility is proportional to the specific heat \cite{QC09,YLG07};
in the case of spin systems described by Hamiltonian of the form (\ref{ham}), with $S$ being a projection of the total spin,
then the fidelity susceptibility is proportional to the initial magnetic susceptibility \cite{V10,QC09,YLG07}. In the non-commutative
case, when the driving term does not commute with the Hamiltonian, such type of relation become more complicated.

It is our goal here to derive lower and upper bounds on theoretical information measures like fidelity susceptibility (\ref{dchi})
and, therefrom, the Bures distance (\ref{infBures}), in terms of thermodynamic quantities like susceptibility
and thermal average values of some
special observables. At that we concentrate on the most interesting non-commutative case.

The paper is organized  as follows: in Section \ref{FS} we introduce our basic notations and derive an
expression for the spectral representation of the fidelity susceptibility in the general
noncommutative case, see (\ref{FiSus2}), which is convenient for the derivation of inequalities involving macroscopic quantities,
like susceptibilities and thermal average values. In  Sections \ref{UB} and  \ref{LB}  new upper, (\ref{FiSUpf}), and lower, (\ref{FiSusLo}), bounds on the fidelity susceptibility are derived. In Section \ref{spin} these bounds are checked against the exact expressions for the simplest case of a single quantum spin in an external magnetic field. We also consider the Kondo model (Section \ref{Kondo}) and the Dicke model (Section \ref{Dicke}) as non-trivial examples for testing our upper and lower bounds, and drawing conclusions about the behavior of the fidelity susceptibility itself. Finally, in Section \ref{D}, we make some comments and compare our results to the inequalities derived in \cite{AASC10} for the Green's function based definition of fidelity susceptibility.

\section{Fidelity susceptibility}\label{FS}

Here we consider the fidelity susceptibility (\ref{dchi})
for the one-parameter family of Gibbs states (\ref{roh}) at the point
$h=0$. To this end we rewrite the density matrix (\ref{roh}) identically as
\begin{equation}
\rho(h):= \rho(0) + h\rho'(0)+ (1/2)h^2\rho''(0)+ r_3(h),
\label{r1dr}
\end{equation}
where
\begin{equation}
r_3(h) := \rho(h) -\rho(0) - h\rho'(0)- (1/2)h^2\rho''(0)
\label{r3}
\end{equation}
is expected to be of the order $O(h^3)$ as $h\rightarrow 0$.
Next we calculate directly the derivatives in Eq. (\ref{r1dr}):
\begin{eqnarray}
&&\rho'(0) = \rho(0)\beta\left[\int_0^1 S(\lambda)
\mathrm{d}\lambda - \langle S\rangle_0\right],\label{ro1} \\
&& \rho''(0)=\rho(0)\beta^2 \left\{\left[\int_0^1 S(\lambda)
\mathrm{d}\lambda -\langle S\rangle_0\right]^2 + \langle
S\rangle_0^2 +\int_0^1 \mathrm{d}\lambda S(\lambda) \int_0^\lambda
\mathrm{d}\lambda' S(\lambda')\right. \nonumber \\&& \left. -
\int_0^1 \mathrm{d}\lambda \int_0^\lambda \mathrm{d}\lambda'
 S(\lambda -\lambda')S(\lambda)- \int_0^1
\mathrm{d}\lambda \langle S(\lambda)S\rangle_0 \right\}
\label{ro2}.
\end{eqnarray}
Here
\begin{equation}
S(\lambda) = \mathrm{e}^{\beta T\lambda} S\mathrm{e}^{-\beta T\lambda},
\end{equation}
and $\langle \dots \rangle_0$ denotes average value with the density
matrix $\rho(0)$. For the further consideration it is important to
note that, in conformity with the normalization conditions
$\mathrm{Tr}\rho(h)= \mathrm{Tr}\rho(0)=1$, we have
\begin{equation}
\mathrm{Tr}\rho'(0)= \mathrm{Tr}\rho''(0)=0.
\label{tr0}
\end{equation}
Our aim is to use the expansion (\ref{r1dr}) to calculate the square root
in the definition of fidelity (\ref{defFidel}) with accuracy $O(h^2)$, as $h\rightarrow 0$,
which is sufficient for obtaining the fidelity susceptibility, see
Eq. (\ref{dchi}).

To introduce our notation and for reader's convenience we derive here
the matrix representation for the fidelity susceptibility in some detail. By using a
slight extension of the method of \cite{SZ03}, we set
\begin{equation}
\sqrt{\rho^{1/2}(0)\rho(h)\rho^{1/2}(0)}=\rho(0) +X +Y +Z,
\label{eq1}
\end{equation}
where $X$, $Y$ and $Z$ are operators proportional to $h$, $h^2$ and $h^3$,
respectively, to be defined below. By squaring Eq.~(\ref{eq1}), we obtain:
\begin{equation}
\rho^{1/2}\rho(h)\rho^{1/2}=\rho X + X\rho +X^2 +\rho Y
+Y\rho + \rho Z + Z\rho +XY +YX +O(h^4). \label{eq2}
\end{equation}
Here and below, for brevity of notation we have set $\rho(0)= \rho$.
Equating terms of the same order in $h$ up to $O(h^3)$ yields a set of three equations:
\begin{eqnarray}
&& h \rho^{1/2} \rho'(0)\rho^{1/2} =\rho X + X\rho,\label{eq3a}\\
&& (1/2)h^2 \rho^{1/2}\rho''(0)\rho^{1/2}=X^2 +\rho Y +Y\rho
\label{eq3b},\\&& \rho^{1/2} r_3(h)\rho^{1/2}= \rho Z + Z\rho +XY +YX .
\label{eq3c}
\end{eqnarray}

To proceed with the calculations in the general case, when the operators
$T$ and $S$ do not commute, we introduce a convenient spectral representation.
To simplify the problem, we assume that the Hermitian operator $T$ has a complete
orthonormal set of eigenstates $|n\rangle$, $T|n\rangle = T_n|n\rangle$, where $n=1,2,\dots $,
with non-degenerate spectrum $\{T_n\}$. In this basis the
zero-field density matrix $\rho(0)$ is diagonal too:
\begin{equation}
\langle m|\rho|n\rangle = \rho_m \delta_{m,n} ,\qquad m,n =1,2,\dots .
\end{equation}
In terms of the matrix elements between eigenstates of the Hamiltonian $T$,
Eq.~(\ref{eq3a}) reads
\begin{equation}
h \rho_m^{1/2}\langle m|\rho'(0)|n\rangle \rho_n^{1/2} = (\rho_m +
\rho_n)\langle m|X|n\rangle, \label{eq4a}
\end{equation}
hence we obtain
\begin{equation}
\langle m|X|n\rangle = h \langle m|\rho'(0)|n\rangle
\frac{\rho_m^{1/2}\rho_n^{1/2}}{\rho_m + \rho_n}. \label{eq5a}
\end{equation}
Similarly, the second equation (\ref{eq3b}) yields
\begin{equation}
(1/2)h^2\rho_m^{1/2}\langle m|\rho''(0)|n\rangle \rho_n^{1/2} =
\langle m|X^2|n\rangle + (\rho_m + \rho_n)\langle m|Y|n\rangle,
\label{eq4b}
\end{equation}
hence
\begin{equation}
\langle m|Y|n\rangle = - \frac{\langle m|X^2|n\rangle}{\rho_m + \rho_n}
 +(1/2)h^2 \langle m|\rho''(0)|n\rangle \frac{\rho_m^{1/2}\rho_n^{1/2}}
{\rho_m + \rho_n}. \label{eq5b}
\end{equation}

Now we turn back to Eq. (\ref{eq1}) and take the trace of both sides:
\begin{equation}
\mathrm{Tr}\sqrt{\rho^{1/2}(0)\rho(h) \rho^{1/2}(0)}=1+
\mathrm{Tr}X +\mathrm{Tr}Y +\mathrm{Tr}Z. \label{trace}
\end{equation}
In contrast to Ref. \cite{SZ03}, no relationship between
$\mathrm{Tr}X$ and $\mathrm{Tr}Y$ appears in our scheme. From Eq. (\ref{eq5a}) we obtain
\begin{equation}
\mathrm{Tr}X = (h/2)\mathrm{Tr}\rho'(0) =0,
\label{trX}
\end{equation}
due to the explicit form of $\rho'(0)$, see (\ref{ro1}). Next, taking into account Eq. (\ref{tr0})
and the expressions (\ref{eq5a}), (\ref{eq5b}) for the matrix elements of $X$ and $Y$,
we calculate
\begin{equation}
\mathrm{Tr} Y = - \sum_{m,n} \frac{|\langle m|X|n\rangle|^2}{2\rho_m}
 +\frac{1}{4}h^2 \mathrm{Tr}\rho''(0) =- \frac{1}{4}h^2\sum_{m,n}
 \frac{|\langle m|\rho'(0)|n\rangle|^2}{\rho_m + \rho_n}. \label{tr4}
\end{equation}

Thus, by substitution of the above results into (\ref{trace}), we derive
with accuracy $O(h^2)$ the following expressions for the fidelity (\ref{defFidel}),
where $\rho_1 =\rho(0)$ and $\rho_2 =\rho(h)$,
\begin{equation}
{\cal F}(\rho(0),\rho(h)) \simeq 1 - \frac{1}{4}h^2\sum_{m,n}
 \frac{|\langle m|\rho'(0)|n\rangle|^2}{\rho_m + \rho_n} , \label{eqf}
\end{equation}
and the squared infinitesimal Bures distance (\ref{infBures}),
\begin{equation}
d_B^2(\rho(0),\rho(h)) \simeq \frac{1}{2}h^2\sum_{m,n}
 \frac{|\langle m|\rho'(0)|n\rangle|^2}{\rho_m + \rho_n}. \label{Buresh2}
\end{equation}
Finally, the definition of  (\ref{dchi}) yields (see \cite{SZ03,ZVG07,AASC10} for a similar  matrix
representation of the quantum fidelity susceptibility)
\begin{equation}
\chi_F(\rho) = \frac{1}{2} \sum_{m,n}
\frac{|\langle m|\rho'(0)|n\rangle|^2}{\rho_m + \rho_n}.
\label{FidSusc}
\end{equation}

Here matrix elements of $\rho'(0)$ are given by
\begin{equation}
\langle n|\rho'(0)|m \rangle = \frac{\rho_n -\rho_m}{T_m -T_n}
\langle n|S|m \rangle, \qquad m\not= n, \label{nond}
\end{equation}
and
\begin{equation}
\langle m|\rho'(0)|m \rangle = \rho_m \beta [\langle m|S|m \rangle
-\langle S\rangle_0],  \label{d}
\end{equation}
where
\begin{equation}
\langle S \rangle_0 :=\sum_{n}\rho_{n}\langle n|S|n \rangle.
\end{equation}

With the aid of the above expressions, the fidelity susceptibility
(\ref{FidSusc}) takes the form
\begin{equation}
\chi_F(\rho) = \frac{1}{2}\sum_{m,n, m\not=n} \left[\frac{\rho_n -\rho_m}{T_m
-T_n}\right]^2 \frac{|\langle m|S|n\rangle|^2}{\rho_m + \rho_n} +
\frac{1}{4}\beta^2 \langle (\delta S^d)^2\rangle_0,
\label{FiSus}
\end{equation}
where $\delta S^d = S^d - \langle S^d\rangle_0$, $S^d$ being the diagonal part of the
operator $S$:
\begin{equation}
S^d :=\sum_{m} \langle m|S|m \rangle |m\rangle \langle m|.
\end{equation}
Hence,
\begin{equation}
\langle (\delta S^d)^2 \rangle_0 :=\sum_m \rho_{m}\langle m|S|m \rangle^2 - \langle S\rangle_0^2.
\end{equation}

Finally, by using the identity
\begin{equation}
\rho_m + \rho_n = (\rho_n - \rho_m)\coth X_{mn},
\end{equation}
where
\begin{equation}
X_{mn} \equiv  \frac{\beta (T_m -T_n)}{2},
\label{def}
\end{equation}
we rewrite Eq. (\ref{FiSus}) in the form:
\begin{equation}
\chi_F(\rho) = \frac{\beta^2}{8}\sum_{m,n, m\not=n} \frac{\rho_n
-\rho_m}{X_{mn}} \frac{|\langle n|S|m \rangle|^2}{X_{mn}\coth
X_{mn}} + \frac{1}{4}\beta^2 \langle (\delta S^d)^2\rangle_0 .
\label{FiSus2}
\end{equation}
Here, the first term in the right-hand side  describes the purely quantum contribution to
the fidelity susceptibility, because it vanishes when the operators $T$ and $S$ comute,
while the second term represents the  ``classical'' contribution.

Now we are ready to derive lower and upper bounds on the fidelity susceptibility
by applying elementary inequalities for $(x\coth x)^{-1}$ to the summand
in the above expression.

\section{Upper bound}\label{UB}

The application of the inequality $(x\coth x)^{-1} \leq 1$ to
Eq. (\ref{FiSus2}) readily gives:

\begin{equation}
\chi_F(\rho) \leq \frac{\beta^2}{8}\sum_{m,n, m\not=n} \frac{\rho_n
-\rho_m}{X_{mn}} |\langle n|S|m \rangle|^2 + \frac{1}{4}\beta^2
\langle (\delta S^d)^2\rangle_0 . \label{FiSusUp}
\end{equation}
By comparing the right-hand side to the expression for the
Bogoliubov-Duhamel inner product (see, e.g.,
\cite{DLS,BT11} and references therein) of the self-adjoint operator
$\delta S$ with itself,
\begin{equation}
(\delta S;\delta S)_0 := \int_0^1 \mathrm{d}\tau \langle \delta
S(\tau) \delta S\rangle_0 =\frac{1}{2}\sum_{m,n, m\not=n}
\frac{\rho_n -\rho_m}{X_{mn}} |\langle n|S|m \rangle|^2 +
\langle (\delta S^d)^2\rangle_0, \label{BD}
\end{equation}
we obtain an upper bound in the transparent form
\begin{equation}
\chi_F(\rho) \leq \frac{\beta^2}{4}(\delta S;\delta S)_0.
\label{FiSUpf}
\end{equation}
Note that the right-hand side of the above inequality is priportional
to the initial  thermodynamic susceptibility:
\begin{equation}
(\delta S; \delta S)_0
= -\frac{N}{\beta}\frac{\partial^2 f[H(h)]}
{\partial^{2} h}\mid_{h=0}:=\frac{N}{\beta}\chi_{N},
\label{hi}
\end{equation}
where $f[H(h)]$ is the free energy density of the system described by the
Hamiltonian (\ref{ham}) and $\chi_{N}$ is the susceptibility with respect
to the field $h$.

\section{Lower bound}\label{LB}

A lower bound follows by applying to the
spectral representation for the fidelity susceptibility (\ref{FiSus2})
the elementary inequality
\begin{equation}
(x\coth x)^{-1} \geq 1 -(1/3)x^2.
\label{1}
\end{equation}
Then, from representation (\ref{FiSus2}), we obtain the following lower bound for the
fidelity susceptibility
\begin{eqnarray}
\chi_F(\rho) &\geq &\frac{\beta^2}{4}(\delta S;\delta S)_0 -
\frac{\beta^2}{24} \sum_{m,n, m\not=n} (\rho_n -\rho_m) X_{mn}
|\langle n|S|m \rangle|^2 \nonumber \\ &=&
\frac{\beta^2}{4}(\delta S;\delta S)_0 - \frac{\beta^3}{48}\langle [[S,T], S]\rangle_0.
\label{FiSusLo}
\end{eqnarray}

\section{The case of a single spin in magnetic field}\label{spin}

To test the quality of the derived upper and lower bounds, we consider the simplest
case of a single spin in external magnetic field subject to a transverse-field perturbation.
By choosing the magnetic field in the reference state along the $z$-axis, and the
transverse perturbation along the $x$-axis, the perturbed Hamiltonian takes the form
\begin{equation}
\beta H(h_1) = -h_1\sigma^x - h_3 \sigma^z , \label{ham1}
\end{equation}
where $\sigma^x$ and $\sigma^z$ are the Pauli matrices.
This choice corresponds to $\beta H(h_1)$ given by Eq. (\ref{ham}) with
\begin{equation}
\beta T =  - h_3 \sigma^z ,\qquad \beta S = \sigma^x .\label{TS}
\end{equation}
The fidelity (\ref{defFidel}) now measures the dissimilarities between the
density matrices
\begin{equation}
\rho(h_1) = \frac{\exp \left(h_1\sigma^x + h_3 \sigma^z\right)}
{2\cosh \sqrt{h_1^2 + h_3^2}}, \label{roful}
\end{equation}
and $\rho(0)$. The average values taken with the density matrix
$\rho(0)$ will be denoted by the symbol $\langle \dots \rangle_0$.
Let $|1\rangle$ and $|-1\rangle$ be the eigenvectors of $\sigma^z$, as
well as of the operator $\beta T = -h_3 \sigma^z$: $\beta T|\pm 1\rangle =
T_{\pm 1}|\pm 1\rangle$, where $T_{\pm 1}= \mp h_3$. Therefore, the density
matrix $\rho(0)$ is diagonal in that basis and its nonzero matrix elements are
\begin{equation}
\rho_{\pm 1} := \langle \pm 1|\rho (0)|\pm 1\rangle  =
\frac{\exp(\pm h_3)}{2\cosh h_3}. \label{ropm}
\end{equation}

Next, for the first derivative (\ref{ro1}) we have the expression
\begin{equation}
\rho'(0) := \left. \frac{\partial}{\partial h_1}\rho(h_1,h_3)\right|_{h_1=0}= \rho \int_0^1
\sigma^x(\lambda) \mathrm{d}\lambda - \langle \sigma^x \rangle_0. \label{ro11}
\end{equation}
Taking into account that $\sigma^x |\pm 1\rangle =|\mp 1\rangle$, hence $\langle \sigma^x
\rangle_0 =0$, we obtain that
\begin{equation}
\langle \pm 1|\rho'(0)|\pm 1\rangle  = 0, \qquad \langle \pm 1|\rho'(0)|\mp 1\rangle  =
\frac{\tanh h_3}{2 h_3}. \label{ro1pm}
\end{equation}

Thus, for the fidelity susceptibility we obtain from Eq. (\ref{FidSusc}) the following result
\begin{equation}
\chi_F(\rho) = \frac{|\langle -1|\rho'(0)|1 \rangle|^2 +
|\langle 1|\rho'(0)|-1 \rangle|^2}{2(\rho_{-1} + \rho_1)}=
\frac{\tanh^2 h_3}{4 h_3^2}.
\label{FidSusc1}
\end{equation}
Evidently, the same result follows from expression (\ref{FiSus}), where the only
nonzero matrix elements are $\langle -1|\beta S|1 \rangle = \langle 1|\beta S|-1 \rangle
=1$.

Let us check now the upper bound (\ref{FiSUpf}). First, by setting in (\ref{BD})
$\beta S =\sigma^x$ and $\beta \delta S =\sigma^x$, we obtain
\begin{equation}
\beta^2(\delta S;\delta S)_0 = \frac{\rho_{-1} -\rho_1}
{T_1 -T_{-1}}|\langle -1|\beta S|1 \rangle|^2 +
\frac{\rho_1 -\rho_{-1}}{T_{-1} -T_1}|\langle 1|\beta S|- 1 \rangle|^2
= \frac{\tanh h_3}{h_3}. \label{BD1}
\end{equation}
Then, we note that $\beta S^d$ = 0, hence
\begin{equation}
\chi_F(\rho) = \frac{\tanh^2 h_3}{4 h_3^2} \leq \frac{\tanh h_3}{4 h_3},
\label{FiSusUp1}
\end{equation}
which is equivalent to the initial inequality $(x\coth x)^{-1} \leq 1$.

Consider now the lower bound (\ref{FiSusLo}). An elementary calculation yields
\begin{equation}
\beta^3[[S,T],S]= - h_3[[\sigma^x ,\sigma^z ],\sigma^x]= 4 h_3 \sigma^z, \label{Lo1}
\end{equation}
hence,
\begin{equation}
\beta^3 \langle [[S,T],S]\rangle_0 = 4 h_3 \tanh h_3, \label{av3}
\end{equation}
and
\begin{equation}
\chi_F(\rho) \geq
\frac{\tanh h_3}{4 h_3}\left(1 - \frac{1}{3}h_3^2\right).
\label{FiSusLo1}
\end{equation}
This lower bound amounts to initial elementary inequality (\ref{1})
and it is always valid but makes sense for $h_3^2 < 3$ only.

\section{Application to the Kondo model}\label{Kondo}

For a long time the Kondo model has been one of the challenging quantum many-body problems in
condensed matter physics. Recent considerations have shown that solid state structures
containing interacting spin systems (the Kondo model provides a specific example of such structures)
are  attractive candidates for quantum information processing \cite{AFOV08}.

In its simplest formulation, the effective Kondo Hamiltonian describes
a single magnetic impurity spin interacting with a band of free electrons in a spatial domain $\Lambda$.
The Hamiltonian has the form:
\begin{equation}
 {\mathcal H}^{K}_{\Lambda} = H_{0}-J(S_1 n^x + S_2 n^y + S_3 n^z ), \label{K1}
\end{equation}
where
$$
 H_{0}=\sum _{k,\sigma}\varepsilon (k) c_{k\sigma}^{\dag}c_{k\sigma},
$$
is the Hamiltonian for the conduction electrons. The fermion operator $c_{k,\sigma}^{\dag}$ $(c_{k,\sigma})$
creates (annihilates) a conduction
electron with momentum k and spin $\sigma$, $J$ is the spin-exchange coupling between
the magnetic impurity spin ${\bf S}=(S_{1},S_{2},S_{3})$, ${\bf S}^2 = s(s + 1)$, located at
the origin ${\bf r}=0$ and the conduction-electron spin densities, ${\bf n}(0)=(n^x, n^y, n^z)$ at ${\bf r}=0$.
In this case the Hilbert space naturally takes the form of a tensor product $\mathfrak{H}\otimes\mathbb{C}^{2s+1}$,
where $\mathfrak{H}$ is the Hilbert space of the free electrons, and
$\exp (-\beta {\mathcal H}^{K}_{\Lambda})$ is of trace class and defines a Gibbs state \cite{H70}.

Let us place the system in a homogeneous external field $h$ in the direction of the z-axis,
\begin{equation}
{\mathcal H}^{K}_{\Lambda}(h) = {\mathcal H}^{K}_{\Lambda}-hS_{3}. \label{K2}
\end{equation}
This choice corresponds to $H(h)$ given by (\ref{ham}) with
\begin{equation}
T:={\mathcal H}^{K}_{\Lambda} ,\qquad  S :=  S_3
\label{Kham}.
\end{equation}
Due to the  rotational invariance of the model Hamiltonian $T$, we have
\begin{equation}
\langle S_{3} \rangle_{0}=0, \quad \langle S_{3}^{2} \rangle_{0}=(1/3)s(s+1),
\end{equation}
where
 $\langle ...\rangle_{0}$ denotes average value with the density
matrix
$$\rho(0):=\frac{e^{-\beta {\mathcal H}^{K}_{\Lambda}}}{\mathrm{Tr} e^{-\beta {\mathcal H}^{K}_{\Lambda}}}.$$
Now, our inequalities (\ref{FiSUpf}) and (\ref{FiSusLo}) yield the following bounds on the fidelity susceptibility.
\begin{equation}
\frac{\beta^2}{4}(S_3; S_3)_0 - \frac{\beta^3}{48}
\langle [[S_3,T], S_3]\rangle_0 \leq \chi_F(\rho)\leq
\frac{\beta^2}{4}(S_3;S_3)_0.
\label{FiSusLoio1}
\end{equation}
Here the double commutator and the Bogoliubov -Duhamel  inner product in (\ref{FiSusLoio1})
 can be estimated in terms of
the model constants in (\ref{K1}) with the aid of the remarkable inequalities obtained by R\"{o}pstorff \cite{R76}:
\begin{equation}
1)\qquad 0\leq \langle [[S_3,T], S_3]\rangle_0\leq \frac{2}{3}J \tanh \beta J,\label{dk}
\end{equation}
and
\begin{equation}
2)\qquad \chi_{C}\frac{1-e^{-\beta\epsilon}}{\beta\epsilon}\leq \beta (S_3;S_3)_{0}\leq \chi_{C},
\label{Ri}
\end{equation}
where $\chi_{C}$ is the Curie susceptibility (the initial susceptibility of a free spin)
\begin{equation}
\chi_{C}=\frac{\beta s(s+1)}{3}\label{CS},
\end{equation}
and the dimensionless quantity
\begin{equation}
\beta\epsilon = \frac{\beta J\tanh \beta J}{2s(s+1)}.
\label{be}
\end{equation}
is introduced.

Let us  use the R\"{o}pstorff's inequalities in combination with our inequalities.
From  (\ref{FiSusLoio1}),  (\ref{dk}) and  (\ref{Ri})  we obtain the following bounds on the fidelity susceptibility
\begin{equation}
\chi_{C} \geq \frac{4}{\beta}\chi_{F}(\rho)\geq \chi_{C}\left[\frac{1-\mathrm{e}^{-\epsilon\beta}}{\epsilon\beta}-
\frac{1}{3}\epsilon\beta \right].\label{nvu}
\end{equation}
These bounds can be readily analyzed and the formula exhibits some interesting properties.
Since by definition $\chi_{F}(\rho)$ is a nonnegative function,  the lowest meaningful bound in the left-hand side of
(\ref{nvu}) is zero.  The upper bound gives the maximal possible value of $\chi_{F}(\rho)$.
Note that $f(x):=(1-e^{-x})/x$ is a strictly monotonically decreasing function of $x=[0,\infty)$,
with $\lim_{x\rightarrow 0}f(x)=1$ and $\lim_{x\rightarrow \infty}f(x)=0$.  As a result the equation
$$
\frac{1-\mathrm{e}^{-x}}{x}-\frac{1}{3}x=0
$$
has a unique solution $x=x^*$.  Therefore, the lower bound in (\ref{nvu}) is nonnegative in the interval
$0\leq \beta\epsilon \leq x^{*}.$

In the limit
$$
\beta\epsilon = \frac{\beta J\tanh \beta J}{2s(s+1)} \rightarrow 0
$$
we obtain an asymptotic relation between the fidelity susceptibility and Curie magnetic susceptibility  (\ref{CS})
\begin{equation}
\chi_{F}(\rho)\simeq\frac{\beta}{4}\chi_{C}.\label{sr}
\end{equation}
From the previous consideration it is known that the above relation becomes equality
if the driving term $S$ and the Hamiltonian $T$ commute, irrespectively of the concrete model.
Indeed, this is the case in the Kondo problem  in the classical limit of infinite spin, $s\rightarrow \infty$,
with $\beta$ held fixed.  Another possibility (\ref{sr}) to take place asymptotically is $\beta J\rightarrow 0$.
Note that, in the case $\infty > \beta\epsilon > x^{*}$, where  the lower  positive bound follows from the definition of $\chi_{F}(\rho),$
we obtain the inequalities
\begin{equation}
0<\chi_{F}(\rho)\leq\frac{\beta}{4}\chi_{C}.
\label{fsn}
\end{equation}
between the fidelity susceptibility and Curie magnetic susceptibility (\ref{CS}).

In general, we see that $(4/\beta)\chi_{F}(\rho)$ can be smaller  than the free spin susceptibility
but exceeds some nonnegative value given by
$$\max \left[\frac{1-e^{-\epsilon\beta}}{\epsilon\beta}-\frac{1}{3}\epsilon\beta ,\quad 0 \right],$$
 which depends on the parameter $\beta\epsilon$.

 If $\beta\epsilon$ is close to zero then $(4/\beta)\chi_{F}(\rho)$ is close to
 the free spin susceptibility. In other words, from the bounds obtained here it is seen that,
when $\beta\epsilon\rightarrow 0$, as well as when $s\rightarrow \infty$,  the
deviation of $(4/\beta)\chi_{F}(\rho)$ from the single-spin Curie law tends to zero.

\section{Application to the Dicke model}\label{Dicke}

The Dicke model \cite{D54} considers the interaction between N two-level atoms and the quantized
single-mode electromagnetic field in a cavity $\Lambda$ of volume $V$. Working at a
 constant density $n =N/V$, one can take the Hamiltonian in the form
\begin{equation}
{\mathcal H}^{D}_N= \omega a^{+}a + \epsilon J^{z}_N +\lambda N^{-1/2}( a + a^{+} )J_N^{x}.
\label{D1}
\end{equation}
Here $a$ and $a^{+}$ denote the annihilation and creation operators
for the cavity mode with frequency $0<\omega<\infty$ which act on the one-mode Fock space $\mathcal{F}_{B}$
and satisfy the commutation relations $[a^+,a^+]= [a^-,a^-]=0, [a^+,a^-]=1$;
 $J_N^{\alpha}=(1/2) \sum_{i=1}^{N}\sigma_{i}^{\alpha}$, where $\sigma^{\alpha}$
$(\alpha=x,y,z)$  are the Pauli matrices acting on a two dimensional complex Hilbert space $\mathcal{C}^{2}$, describe the  ensemble of  two-level atoms, the real $\epsilon$ is the atomic level splitting and
 $0<\lambda<\infty$ is the coupling strength of the atom-field interaction.

With the aid of different methods, it was rigorously proven that in the thermodynamic
limit the Dicke model exhibits a second order phase transition from a normal to a superradiant phase
\cite{HL73a,HL73b,BZT75}.
For values of the parameters $\lambda$, $\omega$  and $\epsilon$  obeying the condition $4\lambda^{2}/\omega
<|\epsilon|$, no phase transition occurs at any temperature, whereas for  $4\lambda^{2}/\omega >|\epsilon|$ there
exists a finite critical temperature $T_{c}=(|\epsilon|/2)\tanh(|\epsilon|\omega/4\lambda^{2})$.
At the point $T=0$, $4\lambda^{2}/\omega =|\epsilon|$, the model exhibits another, quantum phase transition of second order driven by the parameter $\lambda$ .
Quite recently, a renewed interest to this issue appeared due to studies of quantum entanglament.
The superradiant quantum phase transition at the critical point $\lambda_{c}=\sqrt{|\epsilon|\omega}/2$
manifests itself, among other phenomena, in the entanglement between the atomic ensemble and the field mode.

Let us introduce a source term  $S$ for the electromagnetic field in the form
$$
{\mathcal H}^{D}_N(h)={\mathcal H}^{D}_N - hS, \quad S:=\sqrt{N}(a^+ + a)/2,
$$
where $h$ is a real field conjugate to the self-adjoint operator $S$.
Then, the second derivative of the corresponding free energy density with respect to $h$ is
\begin{equation}
\frac{\partial^2 f[{\mathcal H}^{D}_{N}(h)]}
{\partial^{2}h}=-\beta(\delta S; \delta S)_{{\mathcal H}_{N}(h)},
\label{se}
\end{equation}
where $\delta S: = S - \langle S \rangle_{{\mathcal H}^{D}_N(h)}. $
Let us note that  in the thermodynamic limit the second derivative
diverges at the point $(h=0,\quad  T=T_{c})$.

In the case under consideration our inequalities (\ref{FiSUpf}) and  (\ref{FiSusLo}) yield
$$
\frac{\beta^2}{4}(\delta S;\delta S)_0- \frac{\beta^3}{48}\langle
[[S,{{\mathcal H}^{D}_N}], S]\rangle_0\leq
\chi_F(\rho) \leq \frac{\beta^2}{4}(\delta S;\delta S)_0 ,
$$
 where the subscript "0" is used instead of the Hamiltonian ${\mathcal H}^{D}_N$.
One readily calculates
\begin{equation}
\langle[[S,{{\mathcal H}^{D}_N}], S]\rangle_0 = N \omega,
\label{Dcom}
\end{equation}
and, by using the definition (\ref{hi}) of the initial susceptibility per particle with respect to the driving field
$h$, we obtain the bounds
$$
\frac{\beta}{4}\chi_N - \frac{\beta^3 \omega}{48}\leq
\frac{\chi_{F(\rho)}}{N}\leq \frac{\beta}{4}\chi_N
.$$

Using these inequalities one can estimate apart the classical contribution
and the quantum contribution to the fidelity susceptibility. When the size of the system increases $N\rightarrow \infty$  at
the critical temperature $T_{c}$, the thermodynamic susceptibility
$\chi_N\rightarrow \infty$ and for a fixed value of  $\beta \omega/48$ we have the asymptotical result
$$
\frac{\chi_{F(\rho)}}{N}\simeq \frac{\beta}{4}\chi_{N}.$$
Such a relation is a feature of  classical systems. The above relation should be exact if  ${\mathcal H}^{D}_N$
and $S$ commute, which is not the case. However, one can state that due to the strong classical fluctuations
there is a temperature region around $T_{c}$ with borders defined by the condition
$$\beta_{c}^{-1} \chi_N >> \frac{\beta_{c} \omega}{12},$$
where the influence of the quantum fluctuations on the relationship between $\chi_N$ and $\chi_{F(\rho)}/N$ is suppressed.

\section{Discussion}\label{D}

The study of theoretical information properties  of quantum models at the critical point
is a hot topic of a primary interest in the last decade. First, it
allows one to explore the quantum world  as a resource in quantum information processing. Second, it sheds light
on the critical phenomena from a different point of view, being an useful tool in the case
when the standard Landau-Ginzburg approach based on the idea of symmetry breaking is hampered.
As a result, the obtained so far knowledge opens an exciting possibility of the study in a unified framework
of quantum information theory,  critical phenomena paradigms, and novel quantum phases of matter.
Our consideration is an additional step in this direction.

Let us mention that important results on the role of critical fluctuations in systems with broken symmetries
have been obtained in the past using  famous inequalities due to Bogoliubov, Mermin-Wagner, Griffits, among others, see, e.g. \cite{G72,DLS,GN01,R}. It should be stressed that a noteworthy property of this inequality approach is that
the obtained results are exact and cannot be inferred from any perturbation theory. The benefits of having exact statements about strongly interacting systems are difficult to overestimate.  A survey of some other inequalities, used to solve
problems arising in the Approximating Hamiltonian Method, has been  recently  given in \cite{BT11}
 along with a number of references.

Inequalities, as a tool for obtaining interesting exact statements, are also wide-spread and traditional in the
information theory, see e.g.\cite{AW03,P03,MPHUZ}.

Our  lower (\ref{FiSusLo}) and upper (\ref{FiSUpf}) bounds indicate that for the detection of phase transitions
between mixed states the fidelity susceptibility per particle $\chi_F/N$ is as efficient as the usual
susceptibility  $\chi$. This conclusion is in conformity with the commonly accepted view that
quantum fluctuations are dominated by the thermal ones when $T_c >0$. However, one should keep in mind
that our results were derived under rather restrictive conditions on the spectrum of the Hamiltonian.

The same conclusion has been drawn in \cite{AASC10}, although for a different definition of the
fidelity susceptibility. A number of authors, see e.g. \cite{Gu10,AASC10}, start from the expansion of the zero-temperature definition of fidelity (at $\beta =\infty$)
\begin{equation}
| \langle \psi_{0}(h)||\psi_{0}(0)\rangle| \simeq 1 - \frac{1}{2}h^{2}\chi_{F}(\infty),
\quad h\rightarrow 0,
\end{equation}
where $\psi_{0}(h)$ is the ground state eigenfunction of Hamiltonian (\ref{ham}).
In terms of the spectral representation of the operator $H(0) =T$, this yields \cite{ZGC07,YLG07}:
\begin{equation}
\chi_{F}(\infty) =\sum _{n\not=0}\frac{|\langle\psi_{n}(0)|S|\psi_{0}(0)\rangle|^{2}}{[T_{n} - T_{0}]^{2}}.
\label{FSO}
\end{equation}
This quantity is related to the imaginary-time correlation function
\begin{equation}
G(S|\tau)=\theta(\tau)[\langle S(\tau)S\rangle_0 - \langle S \rangle_0^{2}],
\label{dG}
\end{equation}
where $\theta$ is the Heaviside step function, as follows \cite{ZGC07,YLG07}:
\begin{equation}
\chi_{F}^G(\infty)=\int_{0}^{\infty}\tau G(S|\tau) \mathrm{d}\tau.
\label{si0}
\end{equation}
Its finite-temperature generalization  is  defined as \cite{SAC09,AASC10}
\begin{equation}
\chi_{F}^G(\beta)=\int_{0}^{\beta/2}\tau G(S|\tau) \mathrm{d}\tau,
\label{siT}
\end{equation}
 where $G(S|\tau)$ is again given by (\ref{dG}) but the symbol $\langle ...\rangle_0$ now  denotes the thermal average in the Gibbs ensemble with  Hamiltonian $H(0)=T$. A subtle point in this definition is the upper integration limit $\beta/2$, in lieu of $\beta$, which  was justified  by path-integral arguments and a Quantum Monte Carlo method  in \cite{AASC10}.

In Ref.  \cite{S10} it was  found that  definition  (\ref{siT}) of the fidelity susceptibility $\chi_{F}^G(\beta)$ is based on
a definition of the fidelity itself,
\begin{equation}
F^G(\rho_1,\rho_2) = \mathrm{Tr}\sqrt{\rho_1^{1/2} \rho_2^{1/2}}, \label{GFidel}
\end{equation}
which is different from Uhlmann's definition (\ref{defFidel}) that was accepted in the present study.

Let us consider the  bounds on $\chi_{F}^G(\beta)$  obtained in \cite{AASC10}:
\begin{equation}
 \frac{1}{2}ds^2(0,\beta) \leq \chi_{F}^G(\beta)\leq ds^2(0,\beta),
\label{bFG}
\end{equation}
where (in our notation)
\begin{equation}
ds^2(0,\beta)   = \frac{1}{4}\beta^2 \langle (\delta S^d)^2\rangle_0 +
\sum_{n>m} \frac{|\langle m|S|n\rangle|^2}{(T_m -T_n)^2}\;
\rho_n \frac{\left(1 - \mathrm{e}^{-2X_{mn}}\right)^2}{1 + \mathrm{e}^{-2X_{mn}}}
\label{specds}
\end{equation}
Here $ds^2(0,\beta)h^2$ is the leading-order term in the expansion of the squared Bures distance between the states
 $\rho(0)$ and  $\rho(h)$ as $h\rightarrow 0$. As expected from  relation (\ref{infBures}), expression
(\ref{specds}) exactly equals our fidelity susceptibility $\chi_F(\rho)$, as given by spectral representation (\ref{FiSus}).
Thus, inequalities (\ref{bFG}) provide a new lower bound on $\chi_F(\rho)$,
\begin{equation}
\chi_F(\rho) \geq \chi_{F}^G(\beta),
\label{newLB}
\end{equation}
which is a kind of dynamical structural factor of the driving Hamiltonian $S$. An advantage of this bound is
that it can be computed with the aid of the quantum Monte Carlo stochastic series expansion \cite{SAC09}.
From the spectral representation of   \cite{AASC10}
\begin{equation}
 \chi_F(\rho) \geq \chi_{F}^G(\beta)
  = \frac{1}{8}\beta^2 \langle (\delta S^d)^2\rangle_0 +
\sum_{n>m} \frac{|\langle m|S|n\rangle|^2}{(T_m -T_n)^2}\,
 \left(\sqrt{\rho_m} - \sqrt{\rho_n}\right)^2 ,
\label{specLB}
\end{equation}
we conclude that the lower bound (\ref{newLB}) follows from expression (\ref{FiSus}) and the elementary
inequality $$(\rho_n -\rho_m)^2 \geq (\rho_n +\rho_m)(\sqrt{\rho_m} - \sqrt{\rho_n})^2.$$  At that, the
resulting bound will be with the better coefficient 1/4 of the classical term instead of 1/8 as in (\ref{specLB}).

We have established bounds on the fidelity susceptibility which are expressed in terms of thermodynamic
quantities. An additional reasoning that stimulates this line of consideration of
 information-theoretic quantities, is that the experimental setup for measuring
 thermodynamic quantities is well developed. Thus,  estimation of  metric
 quantities with a thermodynamic-based experiment seems to be very appealing.
We have shown that as far as divergent behavior  in the thermodynamic limit is considered, the fidelity
susceptibility $\chi_F $ and the  usual thermodynamic susceptibility $\chi$
are equivalent for a large class of models exhibiting critical behavior.
It remains for the future to study the effect of degeneracy, especially macroscopic one, of the ground state
on the upper and lower bounds for the fidelity susceptibility.

\section*{Acknowledgements}

N.S.T. gratefully acknowledges the support by the National Science Foundation of Bulgaria
under grant TK-X-1712/2007.

\end{document}